\documentclass[english,jfm]{jfm}
\usepackage[T1]{fontenc}
\usepackage[latin9]{inputenc}
\usepackage{verbatim}
\usepackage{amstext}
\usepackage{graphicx}
\usepackage{amssymb}
\usepackage[authoryear]{natbib}

\begin{document}


\title{Bottom-trapped currents as statistical equilibrium states above topographic anomalies}

\author[A. Venaille]{A.\ns V\ls E\ls N\ls A\ls I\ls L\ls L\ls E%
  \thanks{antoine.venaille@ens-lyon.org}}

\affiliation{Laboratoire de Physique, ENS-Lyon, 69007 France.}

\date{\today}
\maketitle

\begin{abstract}
Oceanic geostrophic turbulence is mostly forced at the surface, yet strong bottom-trapped flows are commonly observed along topographic anomalies. Here we consider the case of a freely evolving, initially surface-intensified velocity field above a topographic bump, and show that the self-organization into a bottom-trapped current can result from its turbulent dynamics.  Using equilibrium statistical mechanics, we explain this phenomenon as the most probable outcome of turbulent stirring. We compute explicitly a class of solutions characterized by a linear relation between potential vorticity and streamfunction, and predict when the bottom intensification is expected. Using direct numerical simulations, we provide an illustration of this phenomenon that agrees qualitatively with theory, although the ergodicity hypothesis is not strictly fulfilled.
\end{abstract}

\section{Introduction}
Bottom-intensified flows  are commonly observed along topographic anomalies in the ocean. A striking example is given by the Zapiola anticyclone, a strong recirculation about $500$ km wide that takes place above a sedimentary bump in the Argentine Basin, where bottom-intensified velocities of order $0.1$
m.s$^{-1}$ have been reported from \emph{in situ} measurements \citep{SaundersKing95} and models \citep{deMiranda99}. This bottom intensification may seem surprising, since the energy is mostly injected at the surface of the oceans \citep{FerrariWunsch2009}. Indeed, the primary source of geostrophic turbulence is mostly baroclinic instability, extracting turbulent energy from the potential energy reservoir set at {the} basin scale by large scale wind patterns \citep{GGS74}, and involving surface-intensified unstable modes, see e.g. \citep{SmithJMR07,Tulloch10,AVGVSS_JPO10}. It leads to the following question: how does a turbulent flow driven by surface forcing evolve into a bottom-intensified flow? A first necessary step to tackle this problem is to determine if an initially surface-intensified flow may evolve towards a bottom-intensified current without forcing and dissipation. Here we address this issue in the framework of freely-evolving stratified quasi-geostrophic turbulence, which allows for theoretical analysis with equilibrium statistical mechanics.

The main idea of statistical mechanics is to predict the large scale
flow structure as the most probable outcome of turbulent stirring
\citep{MillerWeichmanCross92,RobertSommeria92}. The interest of the
approach lies in the possibility to compute this large scale flow structure
and to study its properties with respect to a few key parameters given
by dynamical invariants, see e.g. \citep{MajdaWangBook,BouchetVenaillePhysRep11}.
We focus here on a particular class of equilibrium states amenable
to analytical treatment, namely, those characterized by a linear relation
between potential vorticity (PV) and streamfunction.%

Phenomenological arguments for the formation of bottom-trapped flows
were previously given by \citet{Dewar98} in a forced-dissipated
case. We provide here a complementary point of view, built upon previous
results of \citet{Merryfield98JFM}, who computed critical states of equilibrium statistical mechanics for truncated dynamics. He observed that some of these states were bottom intensified in {the} presence of topography.  \citet{AVGVSG11} showed how to find the equilibrium states among these critical states, and how they depend on the initial fine-grained enstrophy profile. However, they focused on the role of {the} beta effect in barotropization processes and did not
discuss the effect of bottom topography. {Here we show} that bottom-trapped currents are actual statistical equilibria in the presence of sufficiently large bottom topography, and we present direct numerical simulations of the free evolution of an initial surface-intensified flow towards these bottom-trapped currents.

\section{Equilibrium states of quasi-geostrophic flows}

Here we consider an initial value problem for a  Boussinesq, continuously stratified quasi-geostrophic fluid. Such flows are stably stratified with a prescribed buoyancy profile $N(z)$ above a topography anomaly $h_b(x,y)$, and are strongly rotating at a rate $f_0/2$ such that $UL/f_0 \ll 1$, where $U$ and $L$ respectively are typical horizontal velocity and length of the flow. At leading order, the flow is at geostrophic balance: horizontal pressure gradients compensate the Coriolis term. The dynamics is then given by quasi-geostrophic equations  (see \citet{VallisBook} section 5.4):
\begin{equation}
\partial_{t}q+\partial_{x}\psi\partial_{y}q-\partial_{y}\psi\partial_{x}q=0\ ,\label{eq:FullDynamics}
\end{equation}
\begin{equation}
q=\Delta\psi+\frac{\partial}{\partial z}\left(\frac{f_{0}^{2}}{N^{2}}\frac{\partial}{\partial z}\psi\right)+\beta y,\ \label{eq:PV_definition}
\end{equation}
\begin{equation}
\text{with } \partial_{z} \psi|_{z=0} = 0,\ \ \frac{f_{0}}{N^{2}}\partial_{z}\psi \bigg|_{z=-H}=-h_{b}\ ,
\label{eq:boundary_conditions}
\end{equation}
where $\psi$ is the streamfunction, $\Delta$ the horizontal Laplacian, $H$ the ocean depth. We consider Cartesian coordinates, but the earth sphericity is taken into account at lowest order through the term $\beta y$, which is due to the meridional variations of the projection of the earth rotation rate on the local vertical axis. This  is the beta plane approximation. The lateral buoyancy variations have been neglected in the upper and lower boundary conditions. Quasi-geostrophic approximation neglects  topography variations to leading order. This is why the term $h_b$ appears only at the boundary $z=-H$.

The dynamics (\ref{eq:FullDynamics}-\ref{eq:PV_definition}-\ref{eq:boundary_conditions}) takes place in three dimensions, but is expressed as the advection of a scalar field (the PV $q$) by a non-divergent horizontal velocity field, as in two-dimensional Euler dynamics. For this reason, the phenomenology of stratified quasi-geostrophic turbulence is similar to two-dimensional turbulence: there is an inverse energy cascade, self-organization at the domain scale, and no dissipative anomaly.

Quasi-geostrophic dynamics develops complex PV filaments at finer and finer scales at each depth $z$, associated with a forward cascade of enstrophy \citep{Kraichnan_Motgommery_1980_Reports_Progress_Physics}. Rather than describing the fine-grained structures, equilibrium statistical theories of two-dimensional turbulent flows, by assuming ergodicity (or at least sufficient mixing in phase space, corresponding to stirring in physical space), predict self-organization of the flow on a coarse-grained level (\citet{RobertSommeria92,MillerWeichmanCross92}, MRS hereafter): a {}``mixing entropy'' is maximized by taking into account the constraints, namely, all the flow invariants, which are the total energy
\begin{equation}
\mathcal{E}=\frac{1}{2}\int_{-H}^{0}\mathrm{d}z\ \int_{\mathcal{D}}\mathrm{d}x\mathrm{d}y\ \left[\left(\nabla\psi\right)^{2}+\frac{f_{0}^{2}}{N^{2}}\left(\partial_{z}\psi\right)^{2}\right], \label{eq:energy_psi}
\end{equation}
and the Casimir functionals $\mathcal{C}_{g}(z)[q]=\int_{\mathcal{D}}\mathrm{d}x\mathrm{d}y\ g(q)$ where $g$ is any continuous function, and $\mathcal{D}$ the domain where the flow takes place. Here we consider the case of a doubly-periodic domain. In the case $g(q)=q^2 / 2$, $\mathcal{C}_{g}$ is the enstrophy. The output of the theory is a coarse-grained PV field $\overline{q}$, obtained by locally averaging the fine-grained PV field $q$.

Let us call $E_0=\mathcal{E}(q_0)$ and $Z_{0}(z)=(1/2)\int_{\mathcal{D}}\mathrm{d}x\mathrm{d}y\ q_0^{2}$ the energy and fine-grained enstrophy, respectively, of the initial condition given by $q_0(x,y,z)$. Using a general result of \citet{Bouchet:2008_Physica_D}, it is shown in \citet{AVGVSG11}, Appendix 1, that in the \textit{low energy limit}, the calculation of MRS equilibrium states amounts to finding the minimizer $\overline{q}_{min}$ of the ``total coarse-grained enstrophy''
\begin{equation}
\mathcal{Z}_{cg}^{tot}\left[\overline{q}\right]=\frac{1}{2}\int_{-H}^{0}\mathrm{d}z\ \int_{\mathcal{D}}\mathrm{d}x\mathrm{d}y\ \frac{\overline{q}^{2}}{Z_{0}} \label{eq:Enstrophy_coarse-grained}
\end{equation}
among all the fields $\overline{q}$ satisfying the energy constraint
\begin{equation}
\mathcal{E}[\overline{q}]=\frac{1}{2}\int_{\mathcal{D}}\mathrm{d}x\mathrm{d}y\ f_0 h_b \psi|_{z=-H}-\frac{1}{2}\int_{\mathcal{D}}\mathrm{d}x\mathrm{d}y\ \int_{-H}^{0}\mathrm{d}z\ \left(\overline{q}-\beta y\right)\psi =E_0,
\label{eq:MiniMumEnstrophyVP_Energy}
\end{equation}
where Eq. (\ref{eq:MiniMumEnstrophyVP_Energy}) is obtained after integrating Eq. (\ref{eq:energy_psi}) by parts. 
Independent of the statistical mechanics arguments, the variational problem (\ref{eq:Enstrophy_coarse-grained}-\ref{eq:MiniMumEnstrophyVP_Energy}) can be seen as a generalization to the stratified case of the phenomenological minimum enstrophy principle of \citet{BrethertonHaidvogel}. Besides, these states are statistical equilibria of the truncated dynamics \citep{Kraichnan_Motgommery_1980_Reports_Progress_Physics,SalmonHollowayHendershott:1976_JFM_stat_mech_QG} in the limit of infinite wave-number cut-off \citep{Carnevale_Frederiksen_NLstab_statmech_topog_1987JFM}. 

Critical states of the variational problem (\ref{eq:Enstrophy_coarse-grained}-\ref{eq:MiniMumEnstrophyVP_Energy})
are computed by introducing Lagrange multiplier%
\footnote{The inverse temperature $\beta_{t}$ should not be confused with the
{$\beta$} effect. %
} $\beta_{t}$ associated with the energy constraint, and by solving
$\delta\mathcal{Z}_{cg}^{total}+\beta_{t}\delta\mathcal{E}=0,$ where
 variations of the functionals are taken with respect to $\overline{q}$,
 leading to  the linear relation $\overline{q}=\beta_{t}Z_{0}\psi$, see also \cite{AVGVSG11}.  The next step is to find which of these critical states are actual minimizers of the coarse-grained enstrophy for a given energy.

\section{Analytical calculations and simulations}
\begin{figure}
\includegraphics[width=1\textwidth]{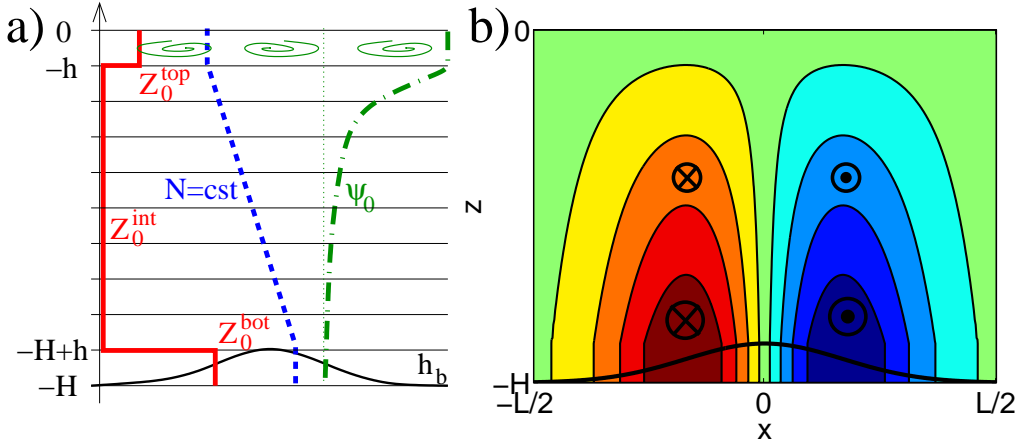}

\caption{a) Sketch of the flow configuration. The continuous red line represents
the initial fine-grained enstrophy profile (here $Z_0^{int}=0$, as assumed in the Appendix). The dashed blue line represents the density profile, and the dashed-dotted green line represents the streamfunction amplitude shape, which is initially surface-intensified. The thick continuous black line represents bottom topography. b) Vertical slice of the meridional velocity field $v$ of the statistical equilibrium state in the low energy limit.\label{fig:sketch} }
\end{figure}

We consider the configuration of Fig. \ref{fig:sketch}-a: the stratification
is linear in the bulk ($N$ is constant for $-H+h<z<-h$), and homogeneous
in two layers of thickness $h\ll H$ at the top and at the bottom,
where $N=0^{+}$. In these upper and lower layers, the streamfunction
is depth independent, denoted by $\psi^{top}(x,y,t)$ {and} $\psi^{bot}(x,y,t)$,
respectively. In these layers, the dynamics is then fully described by the advection of the vertical average of the PV fields, denoted by $q^{top}(x,y,t)$ {and} $q^{bot}(x,y,t)$. The interior PV field is denoted by $q^{int}(x,y,z,t)$. For a given field $q^{top},q^{int},q^{bot}$, the streamfunction is obtained by inverting the following equations: 
\begin{equation}
q^{top}-\beta y =\Delta\psi^{top}-\frac{f_{0}^{2}}{hN^{2}}\frac{\partial}{\partial z}\psi\bigg|_{z=-h}\ ,
\label{eq:invert_q_a}
\end{equation}
\begin{equation}
q^{bot}-\beta y-f_0 \frac{h_{b}}{h} =\Delta\psi^{bot}+\frac{f_{0}^{2}}{hN^{2}}\frac{\partial}{\partial z}\psi\bigg|_{z=h-H}\ , 
\label{eq:invert_q_c}
\end{equation}
\begin{equation}
q^{int}-\beta y=\Delta\psi+\frac{f_{0}^{2}}{N^{2}}\frac{\partial^{2}}{\partial z^{2}}\psi\quad\mbox{for }-H+h<z<-h \ ,
\label{eq:invert_q_b}
\end{equation}
\begin{equation}
\psi^{top}=\psi(x,y,-h),\ \psi^{bot}=\psi(x,y,-H+h).
\label{eq:invert_q_d}
\end{equation}
Equations (\ref{eq:invert_q_a}-\ref{eq:invert_q_c}) are obtained by averaging Eq. (\ref{eq:PV_definition}) in the vertical direction in the upper and the lower layers, respectively, and by using the boundary condition (\ref{eq:boundary_conditions}). In the following, the initial condition is a surface-intensified velocity field induced by a perturbation of the PV field confined in the upper layer: 
\begin{equation}
q_{0}^{top}=\beta y +q_0^{pert}\ , \quad q_{0}^{int}=\beta y\ , \quad q_{0}^{bot}= \frac{f_0}{h} h_b+\beta y.
\end{equation}
It is assumed in the following that $\beta y \ll q_0^{pert} \ll f_0 h_b/h$. The PV fields are therefore associated with fine-grained enstrophies $ Z_0^{int} \ll Z_0^{top} \ll Z_0^{bot}$. 
We compute in the Appendix the coarse-grained enstrophy minimizers of this configuration by solving the variational problem (\ref{eq:Enstrophy_coarse-grained}-\ref{eq:MiniMumEnstrophyVP_Energy}). The main result is that for a fixed topography, in the\emph{ low energy limit},  the  equilibrium streamfunction is a bottom-intensified quasi-geostrophic flow such that bottom streamlines follow contours of topography with positive correlations, see Fig. \ref{fig:sketch}-b. Given the choice of our initial condition, considering the low energy limit for a fixed topography is equivalent to considering a large topography limit for a fixed energy.   

\begin{figure}
\includegraphics[width=1\textwidth]{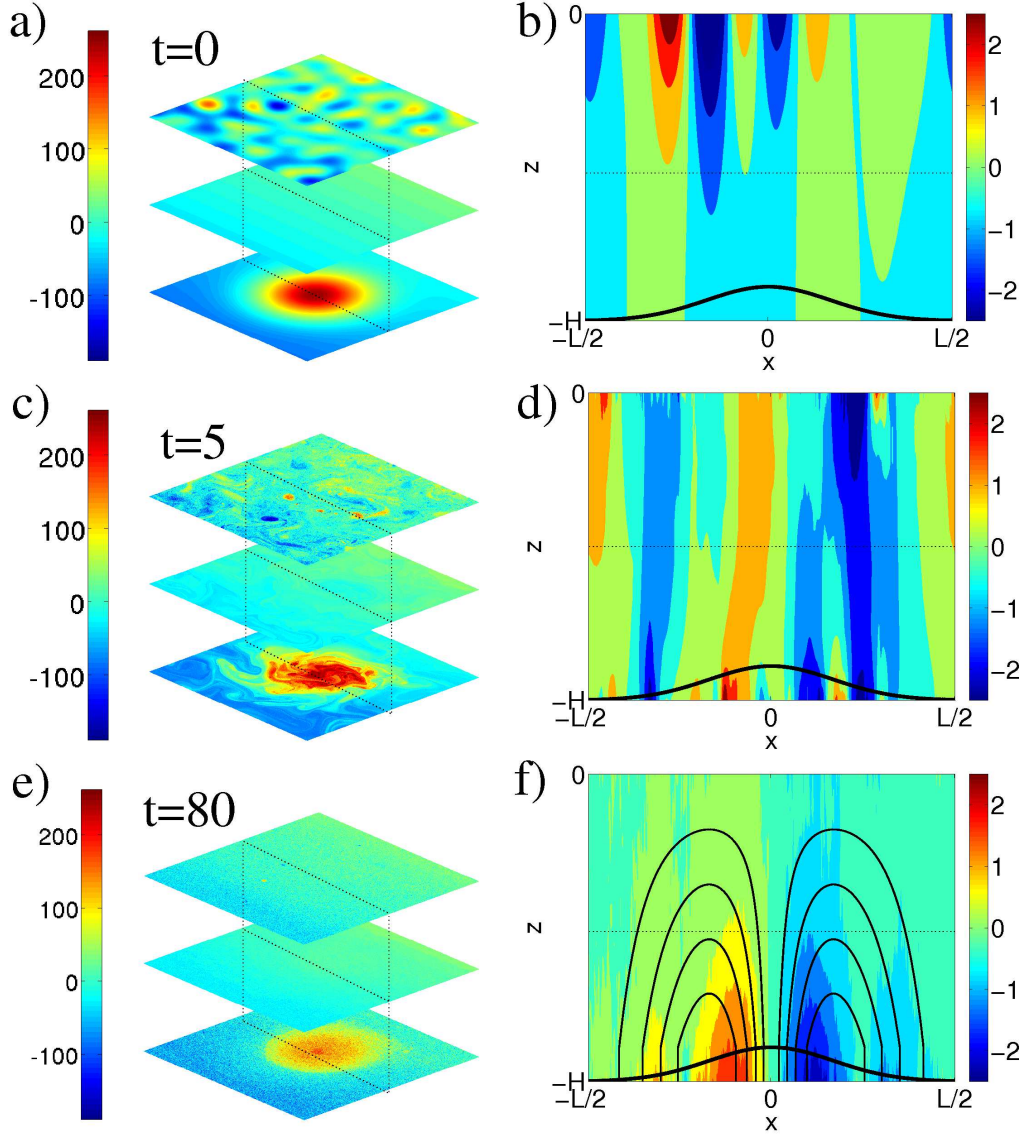}

\caption{a),c),e) PV field at three successive times. In each panel, only layers 1 (top), 5 (middle) and 10 (bottom) are represented. b),d),f) Vertical slices of the meridional velocity fields $v$ taken at the center
of the domain ($y=0$), and associated with the PV fields given in panels a),c),e), respectively. The bold continuous dark line represents bottom topography. The continuous black contours of panel f) give the structure of the statistical equilibrium state in the low energy (or large topography) limit, corresponding to Fig. 1-b.  Contour intervals are the same as those between the different shades.\label{fig:evol} }
\end{figure}

We perform simulations of the dynamics by considering a horizontal discretization of $512^{2}$ in a doubly periodic square domain of size $L=2\pi$, and a vertical discretization with $10$ layers of equal depth $h=0.1$, using a pseudo-spectral quasi-geostrophic model \citep{SmithVallis01}.  The linearly stratified region of Fig. \ref{fig:sketch} is therefore approximated by $8$ layers of depth $h$. We present results without small scale dissipation, but we checked that adding a weak enstrophy filter or hyperviscosity did not change qualitatively the results. 

 The  perturbation $q_{0}^{pert}$ of the PV field in the upper layer has random phases in spectral space with a Gaussian power spectrum peaked at wavenumber $K_{0}=5$ , with variance $\delta K_{0}=2$, and normalized such that the total energy is equal to one ($E_{0}=1$), the latter amounting to a rescaling of time.  The bottom topography $h_{b}$ is a Gaussian bump of amplitude $H_{b}$ and of typical width $\sigma=L/2$. There are three independent parameters: 
\begin{equation}
(i)\ \frac{L_{d}}{L}=\frac{NH}{2f_{0}L},\quad(ii)\ \frac{L_{\beta}}{L}=\frac{1}{L}\left(\frac{V_{0}}{\beta}\right)^{1/2},\quad(iii)\ \frac{L_{h_b}}{L}=\frac{1}{L^{1/2}}\left(\frac{hV_0}{2f_0 H_{B}}\right)^{1/2}, \label{Eq:Paremeters}
\end{equation}
where $V_0=\left(E_0/L^2 H\right)^{1/2}$ is an estimate for the velocity. The first parameter is the ratio between the first baroclinic radius of deformation and the domain scale, taken to be $L_{d}/L=1/10$.  The second one is the ratio between the Rhines scale and the domain scale. We choose $L_{\beta}/L=1/3$.  The third one is the ratio of a topographic Rhines scale to the domain scale, taken to be $L_{h_b}/L=1/10$. We show in the Appendix that the latter choice is consistent with the low energy or large topography limit considered in the previous section. The choices of $\beta$ and $H_b$ are also consistent with the hypothesis that $Z_{0}^{int}\ll Z_{0}^{top} \ll Z_{0}^{bot}$. The choice  $K_0=6$ for the initial condition is motivated by the fact that in the forced-dissipated case, the energy is injected through baroclinic instability with a most unstable mode having a wavelength of a few Rossby wavelength of deformation, see e.g. \citep{SmithJMR07,VallisBook}. We checked that changing this initial horizontal length scale did not change the results.

We see {in} Fig. \ref{fig:evol}-b that the initial condition resembles
a classical surface quasi-geostrophic mode \citep{HeldSQG}. 
After a few eddy turnover times, the enstrophy of the upper layer
has cascaded towards small scales as shown by numerous filaments in Fig.
\ref{fig:evol}-c, {concomitantly} with an increase of the horizontal
energy length scale. The projection of a surface quasi-geostrophic
mode on a Fourier mode of wavenumber $K$ is characterized by a typical
$e$-folding depth of $f_{0}/NK$ on the vertical. The inverse energy cascade on the horizontal leads therefore to a deeper penetration of the velocity field, shown {in} Fig. \ref{fig:evol}-d. When this velocity  field reaches the bottom layer, it starts to stir the bottom PV field. This induces a bottom-intensified flow, which then stirs the surface PV field, and so on. The flow reaches a stationary state after a few dozen of eddy turnover times: the $\overline{q}-\psi$ relation in each layer converges towards a well defined functional relation after coarse-graining. The existence of this functional relation implies $\partial_t\overline{q}=0$, meaning that there are no temporal fluctuations. The corresponding flow is shown in Fig.  \ref{fig:evol}-e, which clearly represents a bottom-intensified anticyclonic flow above the topographic anomaly, qualitatively similar to the one predicted by statistical mechanics in the low energy limit. However, we observe the formation of multiple jets (here 2) around the topographic bump, as in \citet{VallisMaltrud93}, which indicates that the ``sufficient stirring'' hypothesis is not strictly fulfilled in the simulations. 

The same evolution from a surface-intensified turbulent field towards
a bottom-trapped flow was observed in the absence of  a $\beta$-effect, except that there remained two small surface-intensified coherent vortices locked above the topography extrema. These vortices are destroyed by radiation of Rossby waves in presence of $\beta$, as in \citet{Maltrud_Vallis91}. In a way, small values of $\beta$ enhance PV stirring, even if they have a negligible effect on the structure of the equilibrium state. However, if $\beta$ is too large, then  $L_{\beta}\ll L_{h_b}$, which leads to zonal jets that prevent the formation of bottom-trapped flow along topographic anomalies.

\section{Conclusion\label{sec:Conclusion}}
For sufficiently large values of bottom topography, statistical mechanics predicts that an initially surface-intensified velocity field will evolve towards
a bottom-intensified flow following contours of topography.  To the best of our knowledge, this is the first analytical calculation of statistical equilibrium states for stratified quasi-geostrophic equations above topography, although there have been many works on one or few layer models, see e.g. \cite{MajdaWangBook}, and also, critical states were computed by \cite{Merryfield98JFM} in the stratified case. We chose to present these calculations in the simplest possible setting in order to give the main physical ideas. We found qualitative agreement between statistical mechanics predictions and a direct numerical simulation of the freely evolving dynamics. 

The statistical mechanics approach highlights the crucial role of energy and fine-grained enstrophy conservation (among other invariants) while taking  into account the turbulent nature of the dynamics. Here we considered the equilibrium states characterized by energy and enstrophy only, which is justified in a low energy limit, and which leads to linear $\overline{q}-\psi$ relations. The next step will be to consider higher order invariants, in order to account for equilibrium states associated with non-linear $\overline{q}-\psi$ relations, see e.g. \cite{BouchetSimonnetPRL09} in the context of Euler equations. 

The main  shortcoming of the statistical mechanics approach is that it predicts only the final state organization, and not how the system evolves towards this state. In addition, there remain quantitative discrepancies between theory and simulations, because the ergodicity, or ``sufficient stirring'' hypothesis may not be strictly fulfilled in some cases. Addressing specifically the range of parameters for which the statistical mechanics predictions are accurate, and estimating how the relaxation time towards equilibrium depends on these parameters will be the object of a future work.

These results have important consequences for ocean energetics: topographic anomalies allow transferring surface-intensified eddy kinetic energy into bottom-trapped mean kinetic energy, which would eventually  be dissipated in the presence of bottom friction, as for instance  in the case of the Zapiola anticyclone \citep{Dewar98,AVJLSBB_GRL11}. In the case of the Zapiola anticyclone, the dissipation time scale is of the order of a few eddy turnover-time. It is therefore not \textit{a priori} obvious that the results obtained in a freely evolving configuration may apply to this case.  However, we consider that it was a necessary first step. One can now build upon these results to address the role of forcing and dissipation in vertical energy transfers above topographic anomalies.\\

\paragraph{Acknowledgements}
Most of this work was supported by DoE grant DE-SC0005189, by NOAA grant NA08OAR4320752 and by the contract LORIS  (ANR-10-CEXC-010-01). AV warmly thanks F. Bouchet, S. Griffies, I. Held, J. Le Sommer, J. Sommeria and G. Vallis for fruitful discussions and inputs. AV thanks K.S. Smith for sharing his QG code, and J. Laurie and S. Gupta for useful comments on the manuscript. Presentation of the results has been significantly improved thanks to the very detailed comments of an anonymous reviewer.

\section{Appendix: calculation of the solution}

We compute statistical equilibria for the configuration sketched {in} Fig.
\ref{fig:sketch} and detailed in section 3, by applying a general method presented in \citet{AVFB_JSP10}. Here we consider the limit of small beta effect. The case without topography but including the beta effect is discussed in \citet{AVGVSG11}.

When $\beta \rightarrow 0$, only the upper and lower layers are dynamically active. The reason is that the interior fine-grained enstrophy tends to zero. We have seen in section 2 that a solution of the variational problem (\ref{eq:Enstrophy_coarse-grained}-\ref{eq:MiniMumEnstrophyVP_Energy}) necessary satisfies $\overline{q}=\beta_t Z_0 \psi$.  This gives  $\int_{\mathcal{D}}\mathrm{d}x\mathrm{d}y\ \overline{q}^{2}/Z_{0}\sim Z_{0}$. We conclude that $\lim_{Z_{0}\rightarrow0^{+}}\int_{\mathcal{D}}\mathrm{d}x\mathrm{d}y\ \overline{q}^{2}/Z_{0}=0$. The variational problem (\ref{eq:Enstrophy_coarse-grained}-\ref{eq:MiniMumEnstrophyVP_Energy}) amounts therefore to finding the minimizers of
 \begin{equation}
\mathcal{Z}_{cg}^{tot}\left[\overline{q}^{top},\overline{q}^{bot}\right]=\frac{1}{2}\frac{h}{Z_{0}^{top}}\int_{\mathcal{D}}\mathrm{d}x\mathrm{d}y\ \left(\overline{q}^{top}\right)^{2}+\frac{1}{2}\frac{h}{Z_{0}^{bot}}\int_{\mathcal{D}}\mathrm{d}x\mathrm{d}y\ \left(\overline{q}^{bot}\right)^{2},
\label{eq:Zcg_QtopQbot}  
\end{equation}
with the constraint
\begin{equation}
\mathcal{E}[\overline{q}^{top},\overline{q}^{bot}]=-\frac{h}{2}\int_{\mathcal{D}}\mathrm{d}x\mathrm{d}y\ \overline{q}^{top}\psi^{top}-\frac{h}{2}\int_{\mathcal{D}}\mathrm{d}x\mathrm{d}y\ \left(\overline{q}^{bot}-f_0 \frac{h_{b}}{h}\right)\psi^{bot}=E_0.
\label{eq:Energy_QtopQbot}
\end{equation}
One  could also derive this variational problem directly from the Miller-Robert-Sommeria variational problem, using the method presented in the Appendix of \cite{AVGVSG11}, and assuming that PV levels are zero everywhere except in the the lower and upper homogeneous layers.

In order to solve the variational problem (\ref{eq:Zcg_QtopQbot}-\ref{eq:Energy_QtopQbot}), we use the fact that any minimizer
$\overline{q}_{\beta_{t}}^{top},\overline{q}_{\beta_{t}}^{bot}$ of
the "free energy" $\mathcal{F}=\mathcal{Z}_{cg}^{tot}+\beta_{t}\mathcal{E}$
is a minimizer of the enstrophy $\mathcal{Z}_{cg}^{tot}$ with the
energy constraint $E_{\beta_{t}}=\mathcal{E}\left[\overline{q}_{\beta_{t}}^{top},\overline{q}_{\beta_{t}}^{bot}\right]$, see e.g. \citep{AVFB_JSP10} and references therein. 

The first step is to compute the free energy. For that purpose, it is convenient to consider Fourier modes $\psi=\sum_{k,l}\widehat{\psi}_{k,l}exp\left(ikx+ily\right)$ and solve Eq. (\ref{eq:invert_q_b}-\ref{eq:invert_q_d}) for each Fourier component:
\begin{equation}
\widehat{\psi}_{k,l}(z)=\widehat{\psi}_{k,l}^{top}\frac{\sinh\left(z+H/\lambda_{K}\right)}{\sinh\left(H/\lambda_{K}\right)}-\widehat{\psi}_{k,l}^{bot}\frac{\sinh\left(z/\lambda_{K}\right)}{\sinh\left({H}/{\lambda_{K}}\right)}\ ,\quad\lambda_{K}=\frac{f_{0}}{NK}\ ,K=\left(k^{2}+l^{2}\right)^{1/2},
\label{eq:psi_kl}
\end{equation}
where we considered the limit $h\ll H$ to simplify the expressions. Each Fourier component of the streamfunction is a linear combination of a surface quasi-geostrophic mode and a bottom-trapped quasi-geostrophic mode. Substitution of Eq. (\ref{eq:psi_kl}) into the Fourier projections of Eqs. (\ref{eq:invert_q_a}-\ref{eq:invert_q_c}) yields 
\begin{equation}
\widehat{\overline{q}}_{k,l}^{top}=a_{K}\widehat{\psi}_{k,l}^{top}+b_{K}\widehat{\psi}_{k,l}^{bot},\quad\widehat{\overline{q}}_{k,l}^{bot}=b_{K}\widehat{\psi}_{k,l}^{top}+a_{K}\widehat{\psi}_{k,l}^{bot}+\frac{f_0}{h}\left(\widehat{h}_{b}\right)_{k,l},
\label{eq:qtop_qbot_psitop_psibot}
\end{equation}
\begin{equation}
\text{with }a_{K}=-K^{2}\left(1+\frac{\lambda_{K}}{h \tanh\left(H /\lambda_{K}\right)}\right),\quad b_{K}=K^2\frac{\lambda_{K}}{h \sinh\left({H}/{\lambda_{K}}\right)}.
\label{eq:ak_bk}
\end{equation}
In order to compute the free energy $\mathcal{F}=\mathcal{Z}_{cg}^{tot}+\beta_{t}\mathcal{E}$, we first express the enstrophy  (\ref{eq:Zcg_QtopQbot}) and the energy (\ref{eq:Energy_QtopQbot}) in term of the Fourier components of the streamfunction,  potential vorticity fields,  bottom topography. We finally use Eq. (\ref{eq:qtop_qbot_psitop_psibot}) to express the streamfunction in term of the potential vorticity fields and the topography, which yields 
\begin{eqnarray}
\mathcal{F} & =&\frac{h}{2}\sum_{k,l} \left( \left(\frac{1}{Z_{0}^{top}}+\frac{\beta_{t}a_{K}}{a_{K}^{2}-b_{K}^{2}}\right)\left(\widehat{\overline{q}}_{k,l}^{top}\right)^{2}+\left(\frac{1}{Z_{0}^{bot}}+\frac{\beta_{t}a_{K}}{a_{K}^{2}-b_{K}^{2}}\right)\left(\widehat{\overline{q}}_{k,l}^{bot}- \frac{f_0}{h}\left(\widehat{h}_b\right)_{k,l}\right)^{2} \right) \cdots \nonumber\\
& & \cdots + \frac{h}{2}\sum_{k,l} \left( \frac{2\beta_{t}b_{K}}{a_{K}^{2}-b_{K}^{2}}\widehat{\overline{q}}_{k,l}^{top} \left(\widehat{\overline{q}}_{k,l}^{bot} -\frac{f_0}{h} \left(\widehat{h}_b\right)_{k,l} \right) \right).
\label{eq:QuadraticFunctional}
\end{eqnarray}

After some manipulations, we find that the quadratic part of $\mathcal{F}$ is
positive definite for {$\beta_{t}\in \left(\beta_{t}^{min},+\infty \right)$}, where $\beta_{t}^{min}$ is the largest root of $\left(\beta_{t}Z_{0}^{top}-a_{1}\right)\left(\beta_{t}Z_{0}^{bot}-a_{1}\right)=b_{1}^{2}$, implying that there exits a unique  minimizer for each {$\beta_{t}\in \left(\beta_{t}^{min},+\infty \right)$}.
In addition, there is at least one unstable (negative) direction for
$\mathcal{F}$ when $\beta_{t} \in \left(-\infty,\beta_{t}^{min}\right)$, implying 
that there is no minimizer for this range of parameters.

When it exists, the minimizer $\overline{q}_{min}$ of the functional $\mathcal{F}$ can be explicitly computed for each value of $\beta_t$. The first step is to compute the critical points of the functional, which are solutions of $\delta \mathcal{F}=\delta \mathcal{Z}^{tot}_{cg}+\beta_t \delta \mathcal{E}=0$, where variations are taken with respect to the PV fields $\overline{q}^{bot}$, $\overline{q}^{top}$. This leads to $\overline{q}^{i}=\beta_t Z_0^{i} \psi^{i}$ with $i\in\left\{ bot,top\right\}$. Substituting the Fourier projections of these  expressions in Eq. (\ref{eq:qtop_qbot_psitop_psibot}) yields
\begin{equation}
\widehat{\psi}_{k,l}^{top}=\frac{f_0}{h}\frac{b_{K}}{\left(\beta_{t}Z_{0}^{top}-a_{K}\right)\left(\beta_{t}Z_{0}^{bot}-a_{K}\right)-b_{K}^{2}}\left(\widehat{h}_b\right)_{k,l},\quad\widehat{\psi}_{k,l}^{bot}=\frac{\beta_{t}Z_{0}^{top}-a_{K}}{b_{K}}\widehat{\psi}_{k,l}^{top}.
\label{eq:psi_top_kl}
\end{equation}
For a given non-zero topography, the energy tends to zero when $\beta_{t}\rightarrow+\infty$, and tends to infinity when $\beta_{t}\rightarrow\beta_{t}^{min}$. It is therefore possible to reach any admissible energy by varying the inverse temperature between $\beta_{t}^{min}$ and $+\infty$. It thus means that we have found all the energy-enstrophy statistical equilibria.\\

Let us first consider the limit $\beta_{t} Z_0^{bot} L^2 \rightarrow + \infty$. The solution (\ref{eq:psi_top_kl}) tends to
\begin{equation} 
\widehat{\psi}_{k,l}^{top}\approx \frac{f_0 b_{K} \left(\widehat{h}_b\right)_{k,l}}{(h \beta_{t}^{2}Z_{0}^{top}Z_{0}^{bot})},\quad \widehat{\psi}_{k,l}^{bot} \approx \frac{f_0 \left(\widehat{h}_b\right)_{k,l}}{h \beta_{t} Z_{0}^{bot}}\ . \label{eq:lowE}
\end{equation}
We see that $\psi^{bot}=f_0 h_b/(h \beta_t Z_0^{bot})$  is along contours of topography, and that the vertical structure of the equilibrium state is dominated by the bottom-trapped quasi-geostrophic mode, since $\psi^{top}\ll\psi^{bot}$. Using the energy expression (\ref{eq:energy_psi}), we obtain $E_0 \sim H L^2 \left(f_0 H_b /  L h \beta_t Z_0^{bot}\right)^2$, and substituting this expression in  Eq. (\ref{Eq:Paremeters}-iii) yields $\beta_t Z_0 L^2 \sim \left(L/L_{h_b}\right)^2$. These estimates explain \textit{a posteriori} the term \textit{low energy limit} or \textit{large topography limit} when $\beta_t Z_0^{bot}L^2\rightarrow+\infty$, and the fact that this limit corresponds to $L_{h_b}\ll L$.\\

In the \textit{low topography limit or large energy limit}, the ratio $Z_0^{bot}/Z_0^{top}$ tends to zero, and we deduce from Eq. (\ref{eq:psi_top_kl}) that $\beta_{t}\approx\beta_{t}^{min}\approx \left(a_{1}-b_{1}^{2}/a_{1}\right)/Z_{0}^{top}$, which implies that $\psi^{top}$ is dominated by its Fourier component on the horizontal mode $K=1$. We also deduce from  Eq. (\ref{eq:psi_top_kl}) that  $\psi^{bot} \approx-b_{1}/a_{1}\psi^{top}\ll \psi^{top}$: the equilibrium state is a surface quasi-geostrophic mode.\\ 
 
We have thus described a class of equilibrium states varying from a  surface quasi-geostrophic mode condensed in the gravest Laplacian eigenmode (\textit{low topography or large energy limit}) to a bottom-trapped quasi-geostrophic mode following contours of topography (\textit{low energy or large topography limit}). In the framework of MRS theory, only the low energy limit is consistent with  the derivation of the quadratic variational problem (\ref{eq:Zcg_QtopQbot}-\ref{eq:Energy_QtopQbot}).  

Finally, the numerical simulations were performed with ten homogeneous layers, which is a slightly different case from the one studied in this Appendix, with linear interior stratification. However, the present analytical results easily generalize to arbitrary stable interior stratification profiles. We do not expect different qualitative behaviors for the equilibrium states, although the convergence towards equilibrium may depend on stratification properties.

\bibliographystyle{jfm}
\bibliography{AV,barotropization}

\begin{thebibliography}{27}
\expandafter\ifx\csname natexlab\endcsname\relax\def\natexlab#1{#1}\fi

\bibitem[{Bouchet}(2008)]{Bouchet:2008_Physica_D}
{\sc {Bouchet}, F.} 2008 Simpler variational problems for statistical
  equilibria of the 2d euler equation and other systems with long range
  interactions. {\em Physica D Nonlinear Phenomena\/} {\bf 237}, 1976--1981.

\bibitem[{Bouchet} \& {Simonnet}(2009)]{BouchetSimonnetPRL09}
{\sc {Bouchet}, F. \& {Simonnet}, E.} 2009 {Random Changes of Flow Topology in
  Two-Dimensional and Geophysical Turbulence}. {\em Physical Review Letters\/}
  {\bf 102}~(9), 094504.

\bibitem[{Bouchet} \& {Venaille}(2011)]{BouchetVenaillePhysRep11}
{\sc {Bouchet}, F. \& {Venaille}, A.} 2011 {Statistical mechanics of
  two-dimensional and geophysical flows}. {\em Physics Reports (in press,
  arXiv:1110.6245)\/} .

\bibitem[{Bretherton} \& {Haidvogel}(1976)]{BrethertonHaidvogel}
{\sc {Bretherton}, F.~P. \& {Haidvogel}, D.~B.} 1976 {Two-dimensional
  turbulence above topography}. {\em Journal of Fluid Mechanics\/} {\bf 78},
  129--154.

\bibitem[{Carnevale} \&
  {Frederiksen}(1987)]{Carnevale_Frederiksen_NLstab_statmech_topog_1987JFM}
{\sc {Carnevale}, G.~F. \& {Frederiksen}, J.~S.} 1987 {Nonlinear stability and
  statistical mechanics of flow over topography}. {\em Journal of Fluid
  Mechanics\/} {\bf 175}, 157--181.

\bibitem[{de Miranda} {\em et~al.\/}(1999){de Miranda}, {Barnier} \&
  {Dewar}]{deMiranda99}
{\sc {de Miranda}, A.~P., {Barnier}, B. \& {Dewar}, W.~K.} 1999 {On the
  dynamics of the Zapiola Anticyclone}. {\em J. Geophys. Res.\/} {\bf 104},
  21137 -- 21150.

\bibitem[Dewar(1998)]{Dewar98}
{\sc Dewar, W.K.} 1998 Topography and barotropic transport control by bottom
  friction. {\em J. Mar. Res.\/} {\bf 56}, 295--328.

\bibitem[{Ferrari} \& {Wunsch}(2009)]{FerrariWunsch2009}
{\sc {Ferrari}, R. \& {Wunsch}, C.} 2009 {Ocean Circulation Kinetic Energy:
  Reservoirs, Sources, and Sinks}. {\em Annual Review of Fluid Mechanics\/}
  {\bf 41}, 253--282.

\bibitem[{Gill} {\em et~al.\/}(1974){Gill}, {Green} \& {Simmons}]{GGS74}
{\sc {Gill}, A.~E., {Green}, J.~S.~A. \& {Simmons}, A.J.} 1974 {Energy
  partition in the large-scale ocean circulation and the production of
  mid-ocean eddies}. {\em Deep-Sea Research\/} {\bf 21}, 499--528.

\bibitem[{Held} {\em et~al.\/}(1995){Held}, {Pierrehumbert}, {Garner} \&
  {Swanson}]{HeldSQG}
{\sc {Held}, I.~M., {Pierrehumbert}, R.~T., {Garner}, S.~T. \& {Swanson},
  K.~L.} 1995 {Surface quasi-geostrophic dynamics}. {\em Journal of Fluid
  Mechanics\/} {\bf 282}, 1--20.

\bibitem[{Kraichnan} \&
  {Montgomery}(1980)]{Kraichnan_Motgommery_1980_Reports_Progress_Physics}
{\sc {Kraichnan}, R.~H. \& {Montgomery}, D.} 1980 {Two-dimensional turbulence}.
  {\em Reports on Progress in Physics\/} {\bf 43}, 547--619.

\bibitem[{Majda} \& {Wang}(2006)]{MajdaWangBook}
{\sc {Majda}, A. \& {Wang}, X.} 2006 {\em {Nonlinear Dynamics and Statistical
  Theories for Basic Geophysical Flows}\/}.

\bibitem[Maltrud \& Vallis(1991)]{Maltrud_Vallis91}
{\sc Maltrud, M. \& Vallis, G.~K.} 1991 Energy spectra and coherent structures
  in forced two-dimensional and geostrophic turbulence. {\em J. Fluid Mech.\/}
  {\bf 228}, 321--342.

\bibitem[{Merryfield}(1998)]{Merryfield98JFM}
{\sc {Merryfield}, W.~J.} 1998 {Effects of stratification on quasi-geostrophic
  inviscid equilibria}. {\em Journal of Fluid Mechanics\/} {\bf 354}, 345--356.

\bibitem[{Miller} {\em et~al.\/}(1992){Miller}, {Weichman} \&
  {Cross}]{MillerWeichmanCross92}
{\sc {Miller}, J., {Weichman}, P.~B. \& {Cross}, M.~C.} 1992 {Statistical
  mechanics, Euler's equation, and Jupiter's Red Spot}. {\em Phys. Rev. A\/}
  {\bf 45}, 2328--2359.

\bibitem[{Robert} \& {Sommeria}(1992)]{RobertSommeria92}
{\sc {Robert}, R. \& {Sommeria}, J.} 1992 {Relaxation towards a statistical
  equilibrium state in two-dimensional perfect fluid dynamics}. {\em Physical
  Review Letters\/} {\bf 69}, 2776--2779.

\bibitem[{Salmon} {\em et~al.\/}(1976){Salmon}, {Holloway} \&
  {Hendershott}]{SalmonHollowayHendershott:1976_JFM_stat_mech_QG}
{\sc {Salmon}, R., {Holloway}, G. \& {Hendershott}, M.~C.} 1976 {The
  equilibrium statistical mechanics of simple quasi-geostrophic models}. {\em
  Journal of Fluid Mechanics\/} {\bf 75}, 691--703.

\bibitem[{Saunders} \& {King}(1995)]{SaundersKing95}
{\sc {Saunders}, P.~M. \& {King}, B.~A.} 1995 {Bottom Currents Derived from a
  Shipborne ADCP on WOCE Cruise A11 in the South Atlantic}. {\em Journal of
  Physical Oceanography\/} {\bf 25}, 329--347.

\bibitem[{Smith}(2007)]{SmithJMR07}
{\sc {Smith}, K.~S.} 2007 {The geography of linear baroclinic instability in
  Earth's oceans}. {\em Journal of Marine Research\/} {\bf 65}, 655--683.

\bibitem[{Smith} \& {Vallis}(2001)]{SmithVallis01}
{\sc {Smith}, K.~S. \& {Vallis}, G.~K.} 2001 {The Scales and Equilibration of
  Midocean Eddies: Freely Evolving Flow}. {\em Journal of Physical
  Oceanography\/} {\bf 31}, 554--571.

\bibitem[{Tulloch} {\em et~al.\/}(2011){Tulloch}, {Marshall}, {Hill} \&
  {Smith}]{Tulloch10}
{\sc {Tulloch}, R, {Marshall}, J., {Hill}, C. \& {Smith}, K.S.} 2011 {Scales,
  growth rates and spectral fluxes of baroclinic instability in the ocean}.
  {\em Journal of Physical Oceanography\/} {\bf 41}, 1057--1076.

\bibitem[{Vallis}(2006)]{VallisBook}
{\sc {Vallis}, G.~K.} 2006 {\em {Atmospheric and Oceanic Fluid Dynamics}\/}.

\bibitem[{Vallis} \& {Maltrud}(1993)]{VallisMaltrud93}
{\sc {Vallis}, G.~K. \& {Maltrud}, M.~E.} 1993 {Generation of Mean Flows and
  Jets on a Beta Plane and over Topography}. {\em Journal of Physical
  Oceanography\/} {\bf 23}, 1346--1362.

\bibitem[{Venaille} \& {Bouchet}(2011)]{AVFB_JSP10}
{\sc {Venaille}, A. \& {Bouchet}, F.} 2011 {Solvable phase diagrams and
  ensemble inequivalence for two-dimensional and geophysical flows}. {\em
  Journal of Statistical Physics\/} {\bf 143}~(2), 346--380.

\bibitem[{Venaille} {\em et~al.\/}(2011{\natexlab{{\em a\/}}}){Venaille}, {Le
  Sommer}, {Molines} \& {Barnier}]{AVJLSBB_GRL11}
{\sc {Venaille}, A., {Le Sommer}, J., {Molines}, J.M. \& {Barnier}, B.}
  2011{\natexlab{{\em a\/}}} {Stochastic variability of oceanic flows above
  topography anomalies.} {\em Geophysical Research Letters\/} {\bf 38}~(16611).

\bibitem[{Venaille} {\em et~al.\/}(2011{\natexlab{{\em b\/}}}){Venaille},
  {Vallis} \& {Smith}]{AVGVSS_JPO10}
{\sc {Venaille}, A., {Vallis}, G.K. \& {Smith}, K.S.} 2011{\natexlab{{\em
  b\/}}} {Baroclinic turbulence in the ocean: analysis with primitive equation
  and quasi-geostrophic simulations}. {\em Journal of Physical Oceanography\/}
  {\bf 41}~(9), 1605--1623.

\bibitem[{Venaille} {\em et~al.\/}(2011{\natexlab{{\em c\/}}}){Venaille},
  {Vallis} \& {Griffies}]{AVGVSG11}
{\sc {Venaille}, A., {Vallis}, G.~K. \& {Griffies}, S.} 2011{\natexlab{{\em
  c\/}}} {The catalytic role of beta effect in barotropization process}. {\em
  Submitted to Journal of Fluid Mechanics, arXiv:1201.0657\/} .

\end{thebibliography}

\end{document}